\documentclass[a4paper,11pt]{article}
\usepackage{pos}
\usepackage{subfigure}
\usepackage{wrapfig}
\usepackage{xspace}
\usepackage[separate-uncertainty=true]{siunitx}
\usepackage{lipsum}

\newcommand{\gam}{$\gamma$\xspace}
\newcommand{\wld}{Westerlund~1\xspace}
\newcommand{\hess}{H.E.S.S.\xspace}
\newcommand{\hesssrc}{HESS~J1646$-$458\xspace}
\newcommand{\hessforty}{HESS~J1640$-$465\xspace}
\newcommand{\hessfortyone}{HESS~J1641$-$463\xspace}
\newcommand{\psrfortyeight}{PSR~J1648$-$4611\xspace}
\newcommand{\psrfifty}{PSR~J1650$-$4601\xspace}
\newcommand{\cxo}{CXOU~J164710.2$-$455216\xspace}

\DeclareSIUnit\pc{pc}
\DeclareSIUnit\kpc{kpc}
\DeclareSIUnit\erg{erg}  
\DeclareSIUnit\year{yr}

\title{The young massive stellar cluster \wld in \gam rays as seen with \hess}
\ShortTitle{The young massive stellar cluster \wld as seen with \hess}

\author*[a]{Lars Mohrmann}
\author[a]{Andreas Specovius}
\author[b]{Romed Rauth}
\author[c]{Stefan Ohm}
\author[a]{Christopher van Eldik}

\affiliation[a]{Friedrich-Alexander-Universit\"at Erlangen-N\"urnberg, Erlangen Centre for Astroparticle Physics,\\
  Erwin-Rommel-Str. 1, 91058 Erlangen, Germany}

\affiliation[b]{University of Innsbruck, Institute for Astro- and Particle Physics,\\
Technikerstr. 25 / 8th Floor, 6020 Innsbruck, Austria}

\affiliation[c]{DESY, Platanenallee 6, 15738 Zeuthen, Germany}

\forColl{\hess} 
 
\emailAdd{lars.mohrmann@fau.de}

\abstract{Massive stellar clusters have recently been hypothesised as candidates for the acceleration of hadronic cosmic rays up to PeV energies.
Previously, the \hess Collaboration has reported about very extended \gam-ray emission around \wld, a massive young stellar cluster in the Milky Way.
In this contribution we present an updated analysis that employs a new analysis technique and is based on a much larger data set, allowing us to constrain better the morphology and the energy spectrum of the emission.
The analysis technique used is a three-dimensional likelihood analysis, which is especially well suited for largely extended sources.
The origin of the \gam-ray emission will be discussed in light of multi-wavelength observations.
}

\FullConference{
        37$^{\rm{th}}$ International Cosmic Ray Conference (ICRC 2021)\\
		July 12th -- 23rd, 2021\\
		Online -- Berlin, Germany
}

\begin{document}
\maketitle

\section{Introduction}
\wld is a young, massive stellar cluster hosting rich populations of Wolf Rayet stars, OB supergiants and Yellow Hypergiants \cite{Clark2005}.
Its age, total mass, and distance have been estimated to $3.5-5\,\si{Myr}$, $\sim\SI{e5}{M_\odot}$, and $\sim\SI{3.9}{\kpc}$, respectively \cite{Clark2005,Kothes2007}.
However, all of these estimates have been disputed and are subject of ongoing debate (see e.g.\ \cite{Aghakhanloo2020,Beasor2021} and references therein).

Stellar clusters and star forming regions have recently been established as sources of \gam rays at GeV energies \cite{Yang2017,Yang2020,Sun2020,Sun2020a}.
Moreover, massive stellar clusters (and in particular \wld) have been hypothesised as sources of PeV cosmic rays \cite{Bykov2015,Aharonian2019}.
The hypothesis can be tested through \gam-ray observations in the TeV energy band.
We note that, for example, the recent measurement of $>\SI{100}{TeV}$ \gam rays from the Cygnus Cocoon by LHAASO \cite{Cao2021} supports this scenario.
Regarding in particular \wld, the detection of \gam-ray emission with a spectrum extending to energies of $\sim\SI{20}{TeV}$ from its vicinity, labelled \hesssrc, has previously been reported by the High Energy Stereoscopic System (\hess) Collaboration \cite{HESS_Westerlund1_2012}.
Here, we present results from an updated \hess analysis of a much larger data set with about 164~hours of live time (compared to 34~hours in the previous publication).

The analysis of the region surrounding \wld is very challenging for imaging atmospheric Cherenkov telescopes (IACTs) like \hess, both because of the large extent of \hesssrc (diameter of $\sim 2^\circ$ \cite{HESS_Westerlund1_2012}) and due to the presence of other nearby \gam-ray sources, in particular \hessforty \cite{HESS_1640_2014} and \hessfortyone \cite{HESS_1641_2014}.
These circumstances render standard analysis techniques, in which the residual hadronic background is typically estimated from source-free regions in the field of view, close to unapplicable.
In order to address this challenge, we perform a likelihood analysis that employs a 3-dimensional (3D) model for the residual hadronic background in each observation.
While this approach is still relatively new in IACT data analysis, its application to the analysis of \hess data has been explored and verified \cite{Mohrmann2019}.
We carry out the analysis using the Gammapy software package \cite{Deil2017} (version 0.17), which naturally provides the possibility to perform likelihood analyses based on 3D background models.

\section{\hess data set and analyis}
\hess is an array of five IACTs located in the Khomas Highland in Namibia.
The first four telescopes (CT1-4), featuring \SI{12}{\meter}-diameter mirrors, were taken in operation in 2004 and are arranged in a square with \SI{120}{\meter} side length.
A fifth telescope (CT5), with \SI{28}{\meter} mirror diameter, was added in the centre of the array in 2012.
Because a considerable portion of the data set analysed here was recorded prior to the installation of CT5, data from this telescope have not been used for this contribution.

The \wld data set comprises a total of 362 observations of (up to) 28~minutes each, taken between May~2004 and October~2017, adding up to a total live time of 164.2~hours.
Owing to the fact that part of the observations were primarily targeted at the neighbouring sources \hessforty and \hessfortyone, the pointing directions of the observations are not distributed symmetrically around \wld, further adding to the complexity of the analysis.
We reconstruct the incoming direction and energy of \gam-ray candidate events using the algorithm introduced in \cite{Parsons2014} and perform the rejection of hadronic background events with the method detailed in \cite{Ohm2009}.
For each observation, we construct a model for the residual hadronic background, depending on two field-of-view coordinates and reconstructed energy, following the procedure described in \cite{Mohrmann2019}.
We then employ Gammapy to carry out the subsequent high-level analysis.

As a first step in the analysis, we determine a lower energy threshold for each observation, requiring both that the energy bias is tolerable (below 10\%) and that the background model is reliable above this threshold (see \cite{Mohrmann2019} for details).
We furthermore enforce a minimal energy threshold of \SI{0.37}{TeV} in order to ensure that all parts of the observed region of interest have sufficient exposure down to this energy.
To improve the description of the residual background, we then adjust the background model to the observed data for each observation by fitting its normalisation and spectral slope above the energy threshold of the observation.
For this adjustment, we consider only events from within $2^\circ$ of the pointing direction and from outside an exclusion map that covers regions with significant \gam-ray emission.
We obtain the exclusion map through an iterative procedure, where we first adjust the background model using the current map (beginning with an empty one), and generate a new map by identifying regions with an excess of events corresponding to a significance of greater than $4\sigma$ and enlarging these regions by $0.3^\circ$ in all directions.
The algorithm converges after 5~iterations and we find a very good agreement between the background model and the observed data outside the final exclusion map.

\begin{figure}
    \centering
    \subfigure[]{
        \includegraphics[width=0.48\textwidth]{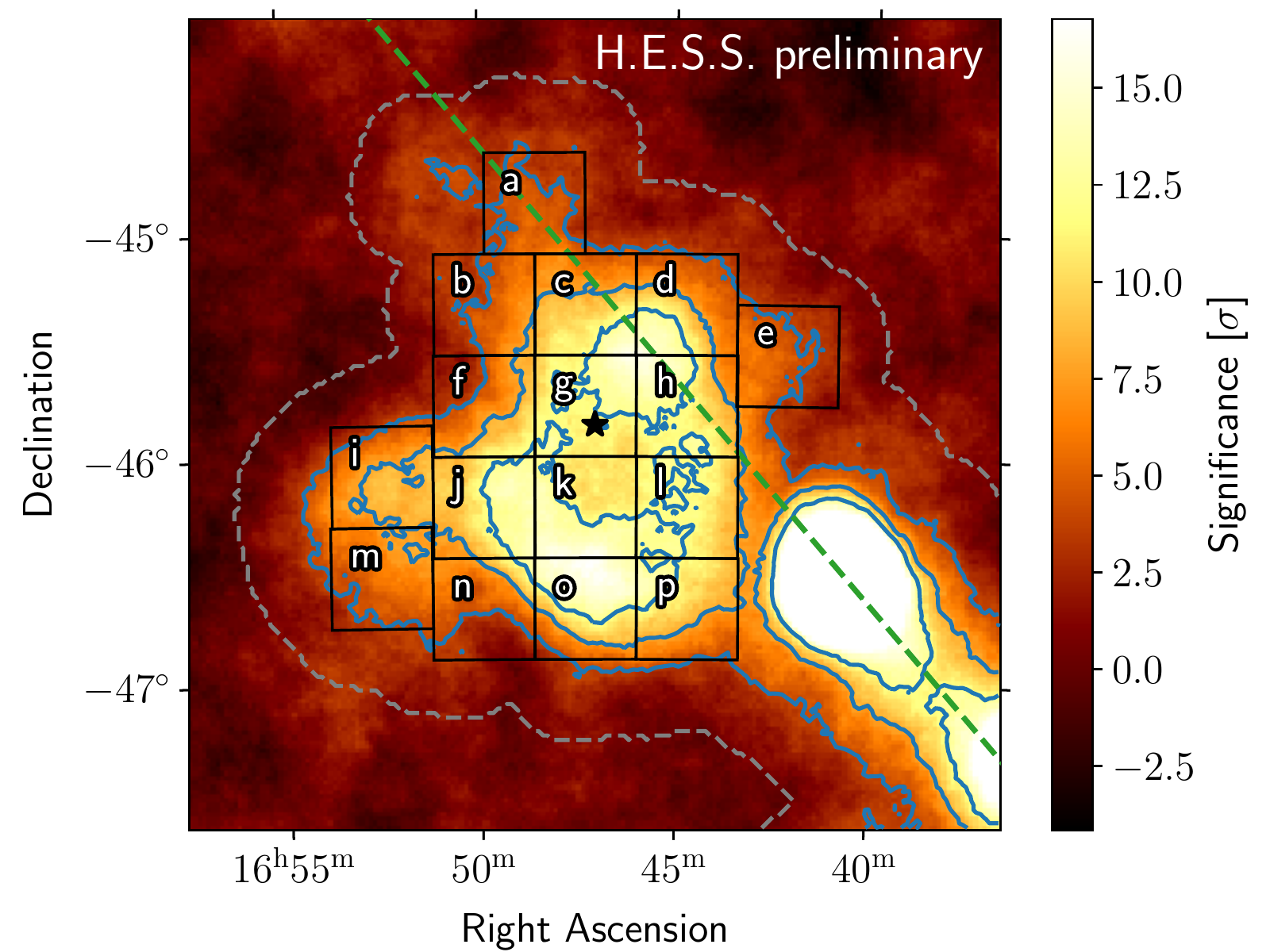}
        \label{fig:sign_map}
    }
    \subfigure[]{
        \includegraphics[width=0.48\textwidth]{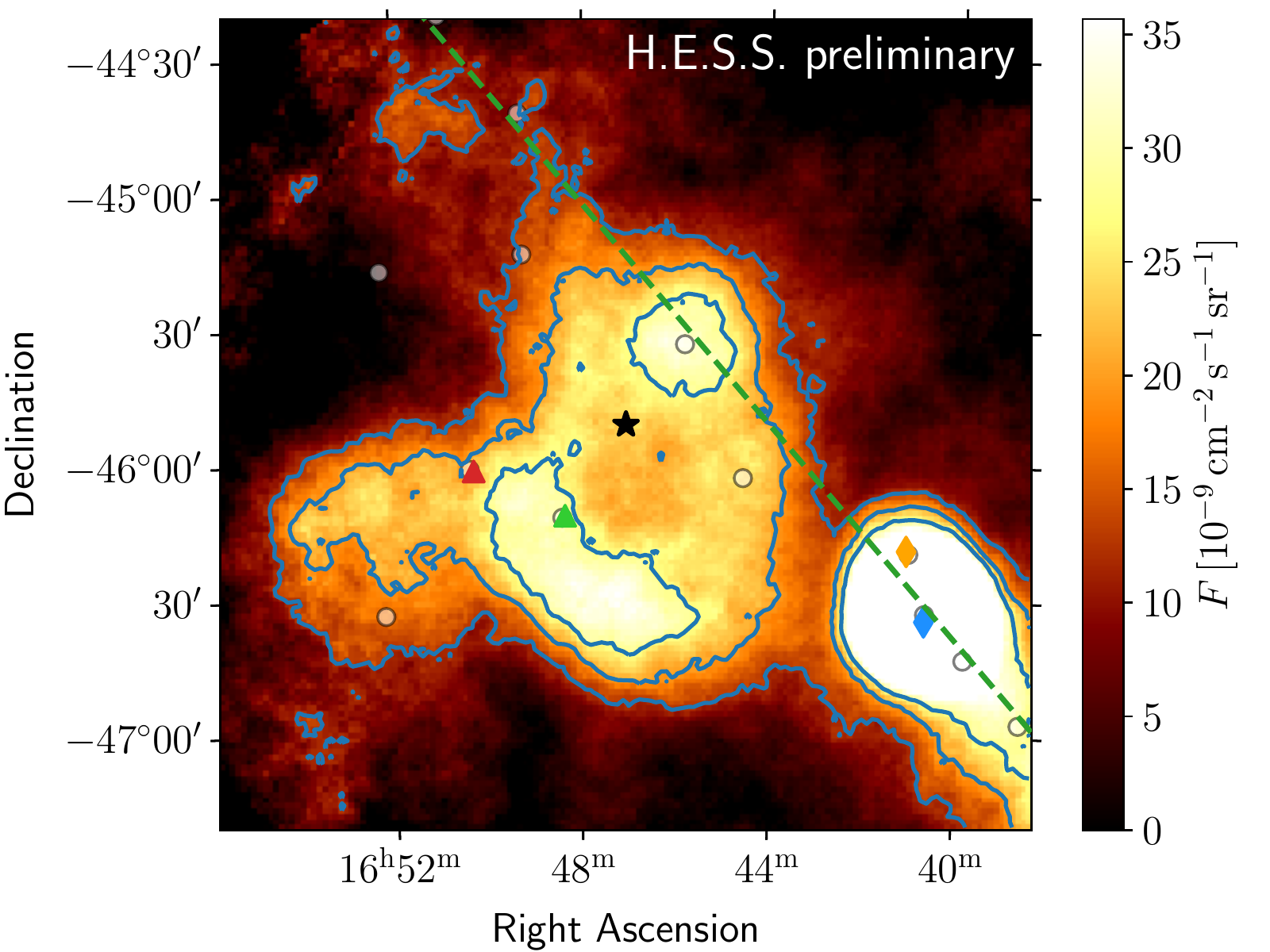}
        \label{fig:flux_map}
    }
    \caption{
        Sky maps of the \wld region, for energies above \SI{0.37}{TeV}.
        The position of the stellar cluster is marked by the black star, the green dashed line denotes the Galactic plane.
        (a) Significance map, with contour lines at 4 / 8 / 12$\sigma$. The grey, dashed line displays the boundary of the exclusion map. Box regions used for the extraction of spectra are indicated as well.
        (b) Flux map, with contour lines at $(12.5 / 20 / 27.5)\times \SI{e-9}{cm^{-2}s^{-1}sr^{-1}}$. The blue and orange diamond mark the position of \hessforty and \hessfortyone, respectively, whereas the green and red triangle show the location of \psrfortyeight and \psrfifty, respectively. Sources from the \textit{Fermi}-LAT 4FGL catalogue \cite{Fermi_4FGL} are marked with circles.
        Note that both color scales are saturated at the maximum observed value for \hesssrc.
    }
    \label{fig:maps}
\end{figure}

\section{Results}
Using the adjusted background model, we derive sky maps that display the \gam-ray emission in the region around \wld, see Fig.~\ref{fig:maps}.
Following \cite{HESS_Westerlund1_2012}, we convolve the measured and predicted counts with a top-hat kernel of $0.22^\circ$ radius before computing the maps, in order to smooth small-scale fluctuations.
Overlaid on the significance map (Fig.~\ref{fig:sign_map}), we display the boundary of the final exclusion map as well as the location of 16 square regions, labelled a--p, that we use in the following to extract energy spectra.
We compute the flux map (Fig.~\ref{fig:flux_map}) assuming that the \gam-ray emission follows a power-law spectrum with index $-2.3$.

Besides detecting a strong \gam-ray signal associated with the nearby sources \hessforty and \hessfortyone, we observe \gam-ray emission from the region surrounding \wld with a very complex morphology.
The emission region, denoted \hesssrc, is largely extended (radius $\sim 1^\circ$) and exhibits various peaks (or ``hot spots'').
Notably, there is no peak at the position of the stellar cluster itself, rather, \wld seems to lie within a region with lower flux, compared to its immediate surroundings.

\begin{wrapfigure}{r}{0.5\textwidth}
    \centering
    \vspace{-0.5cm}
    \includegraphics[width=0.5\textwidth]{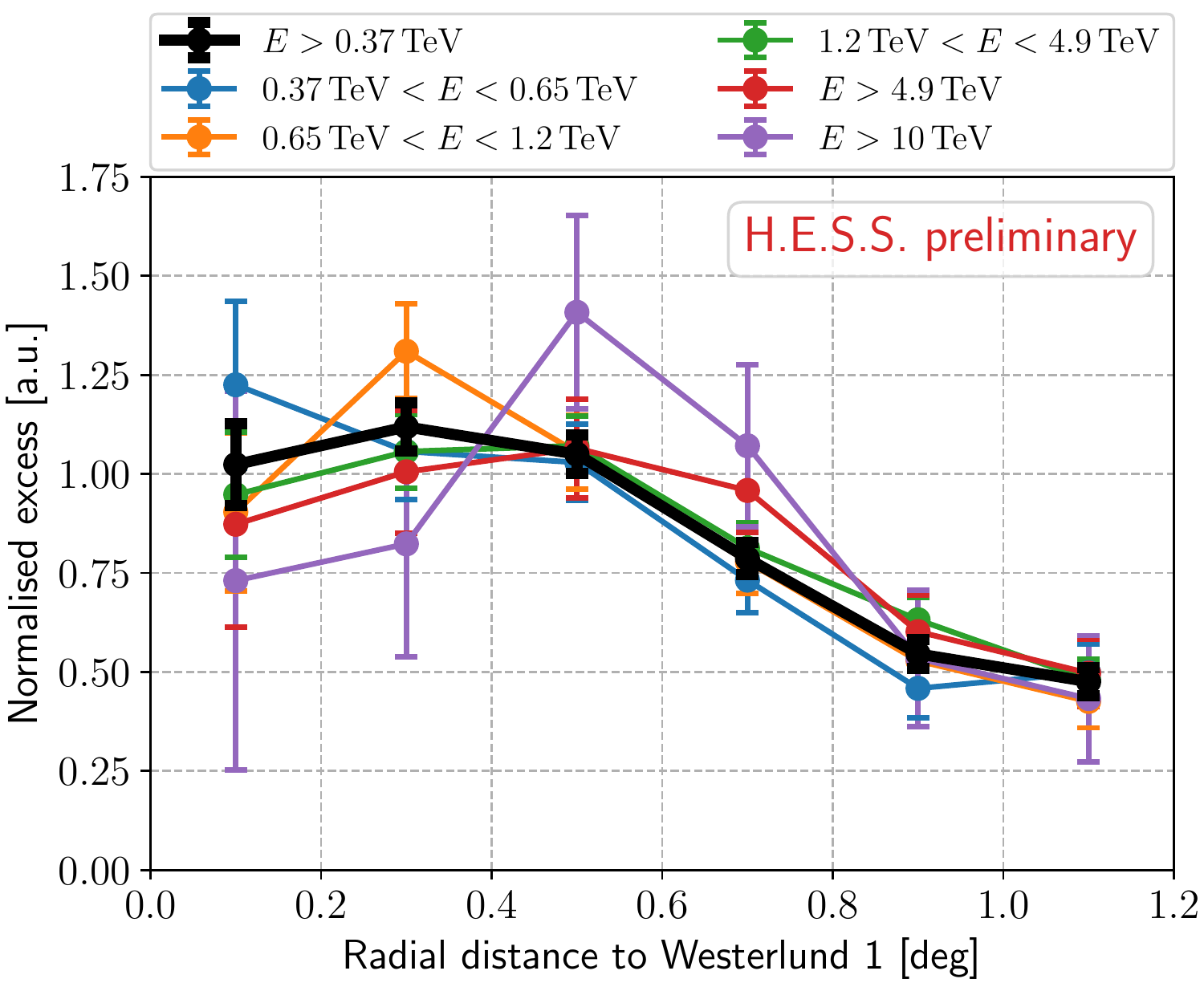}
    \caption{
        Radial profiles of the observed excess events in different energy bands.
        The profiles are acceptance-corrected and normalised, so that their shapes can be compared.
    }
    \label{fig:exc_profile}
\end{wrapfigure}

An important question concerns the observed morphology as a function of energy.
As a test for this, we show in Fig.~\ref{fig:exc_profile} radial profiles of the observed emission, for all energies above the threshold as well as for separate energy bands.
We define the first three energy bands (with boundaries at 0.37, 0.65, 1.2, and \SI{4.9}{TeV}) such that a comparable number of excess events above background is observed in each and consider energies $>\SI{4.9}{TeV}$ and $>\SI{10}{TeV}$ as additional bands.
Table~\ref{tab:ebands} lists all energy bands along with the observed number of excess events within the total of 16 square regions displayed in Fig.~\ref{fig:sign_map}.

We observe no difference in shape between the profiles at different energies.
In order to quantify this finding, we carry out $\chi^2$ tests, comparing the radial profile for each energy band with one that is computed from all events outside this band (such that the two are statistically independent).
The results, listed in the last column of Table~\ref{tab:ebands}, indicate that the radial profiles are consistent with each other across all energies.

\begin{wraptable}{l}{0.5\textwidth}
    \centering
    \vspace{-0.2cm}
    \begin{tabular}{ccc}
        \hline
        \textbf{Energy range [TeV]} & \textbf{Excess} & $\chi^2\,/\,N_\mathrm{dof}$ \\\hline
        $>0.37$ & \num{14169} & $-$\\
        $0.37 - 0.65$ & \num{4852} & 5.43 / 6\\
        $0.65 - 1.2$ & \num{3804} & 4.86 / 6\\
        $1.2 - 4.9$ & \num{4448} & 4.30 / 6\\
        $>4.9$ & \num{1065} & 3.88 / 6\\
        $>10$ & \num{350} & $-$\\
        \hline
    \end{tabular}
    \caption{
        Energy bands considered in the analysis.
        The second column lists the total number of excess events, the third column denotes the result of the shape comparison $\chi^2$ test (cf.\ Fig.~\ref{fig:exc_profile}, see text for details).
    }
    \label{tab:ebands}
    \vspace{-0.7cm}
\end{wraptable}

\begin{figure}
    \centering
    \subfigure[]{
        \includegraphics[width=0.48\textwidth]{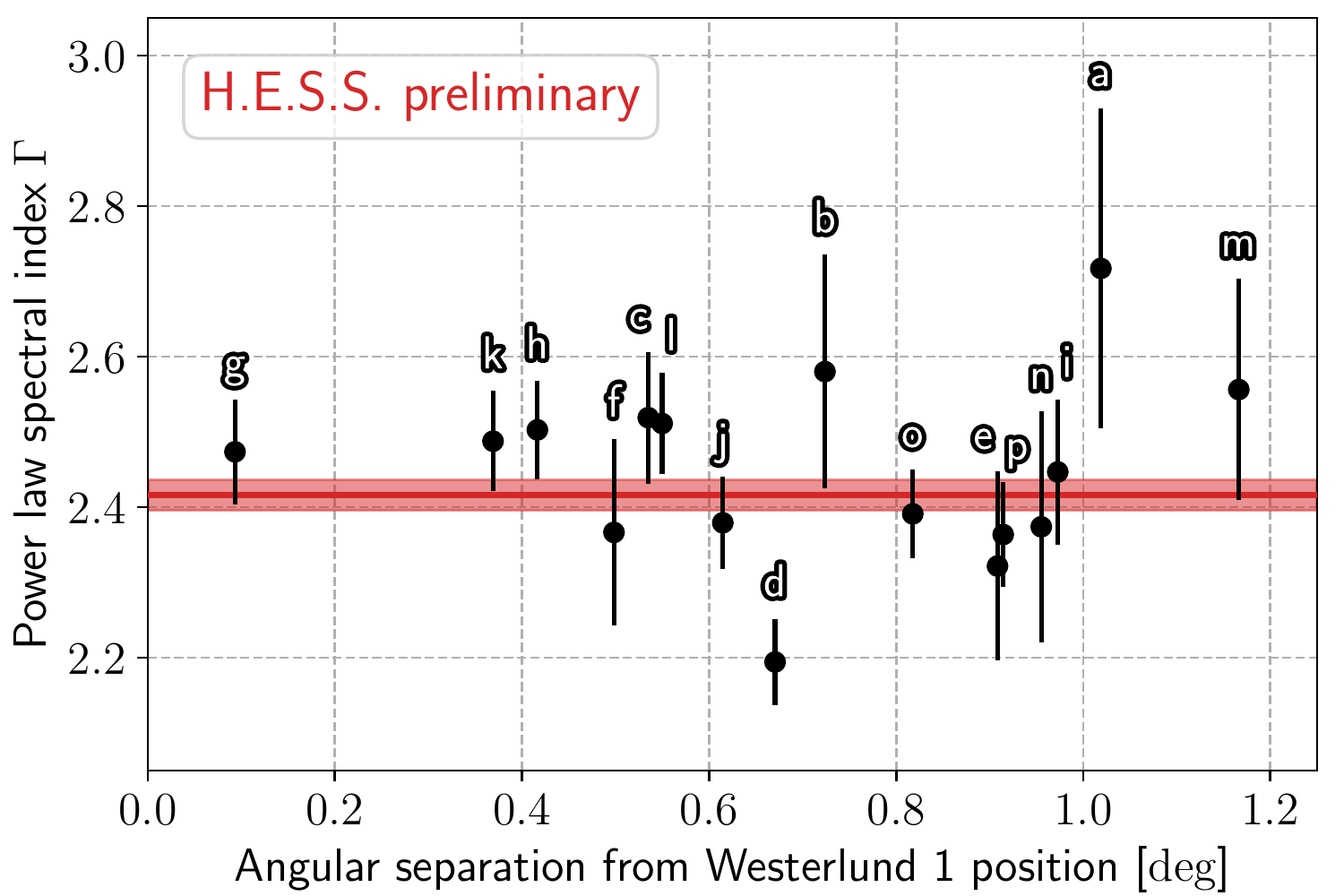}
        \label{fig:box_index_distance}
    }
    \subfigure[]{
        \includegraphics[width=0.48\textwidth]{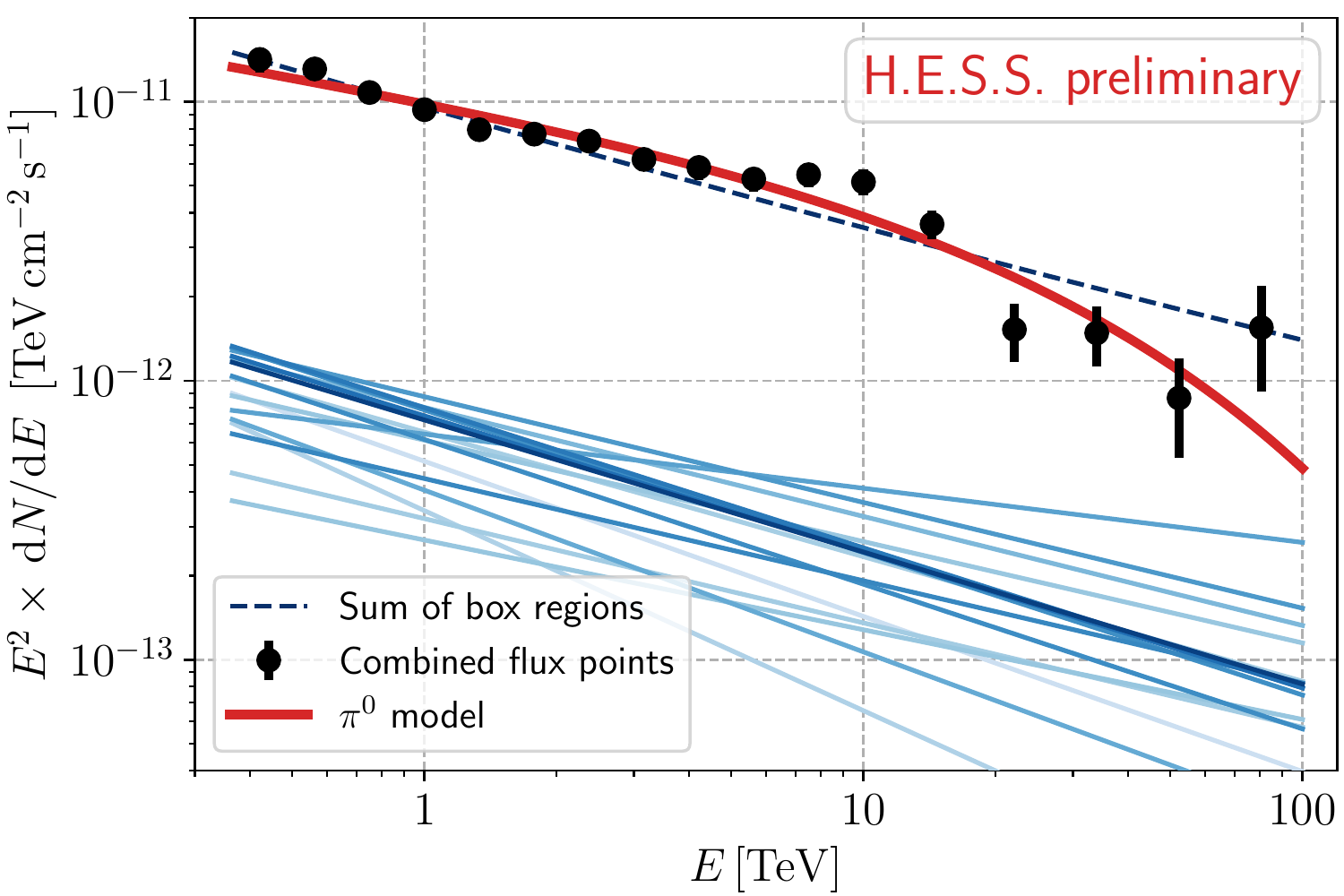}
        \label{fig:combined_spectrum}
    }
    \caption{
        Spectrum results.
        (a) Best-fit spectral index $\Gamma$ of all sub-regions, as a function of the angular separation from the stellar cluster position.
        The red line and band denote the weighted average and its uncertainty, respectively.
        (b) Combined flux points (black) and result of fitting a $\pi^0$-decay model (red line).
        The best-fit power-law spectra of all sub-regions are displayed by the solid blue lines (with darker shades indicating a smaller angular separation of the region from the \wld position); their sum is denoted by the dashed blue line.
        For visualisation purposes we plot all power-law spectra up to \SI{100}{TeV}, however, we note that the observed excess is not significant up to this energy for any of the individual sub-regions.
    }
    \label{fig:spectra}
\end{figure}

In a next step, we aim to study the energy spectrum of the observed emission, both for the entire source region and for sub-regions.
To this end, we extract events (observed and predicted by the background model) from each of the 16 square regions indicated in Fig.~\ref{fig:sign_map} and fit a power-law spectral model ($\mathrm{d}N/\mathrm{d}E \propto E^{-\Gamma}$) to the resulting counts spectra.
In Fig.~\ref{fig:box_index_distance}, we show the best-fit values of $\Gamma$ for each region, as a function of the angular separation of the centre of said region from the position of \wld.
We find that practically all values are compatible with the average index of all regions, shown by the red line in the plot.
The only possible exception is region ``d'', for which the spectral index deviates by about $4\sigma$ from the average of all other regions.
We note, however, that the true spectrum does not need to follow a power law perfectly, and that the deviation may therefore be less significant than suggested by the comparison of the power-law indices.
We conclude that the energy spectrum does not vary in shape significantly across the emission region.

Motivated by the similarity of the energy spectra obtained for the sub-regions, we derive combined flux points for the entire emission region by adding up the flux points of all sub-regions.
These combined flux points are displayed in Fig.~\ref{fig:combined_spectrum}.
In addition to the statistical uncertainty, the error bars contain a systematic uncertainty related to the background model -- contributing mostly at low energies -- that we determine by comparing the rate of events predicted by the model to the observed data in source-free regions positioned outside the exclusion map.

The combined flux points show that the emission from \hesssrc extends to at least several tens of TeV.
While the highest-energy points tend to lie below the expectation from a simple power-law extrapolation, we do not find a clear indication of a cut-off to the spectrum either.
In order to study the possibility that the observed \gam-ray emission is caused by the interaction of cosmic-ray protons with ambient gas, we employ the naima package \cite{Zabalza2015}\footnote{More specifically, we make use of Gammapy's wrapper class for naima models.} to fit a primary proton spectrum ($\mathrm{d}N_p/\mathrm{d}E_p\propto E_p^{-\Gamma_p}\cdot \exp(-E_p/E_p^\mathrm{cut})$) to the combined flux points; the red line in Fig.~\ref{fig:combined_spectrum} denotes the resulting \gam-ray spectrum.
We assume a target gas density of $n=\SI{10}{cm^{-3}}$ and adopt as distance to the source $d=\SI{3.9}{\kpc}$.
The best-fit spectral index and cut-off energy are given by $\Gamma_p=2.33\pm 0.06$ and $E_p^\mathrm{cut}=(400^{+250}_{-130})\,\mathrm{TeV}$, respectively.
For the total energy in protons with $E_p>\SI{1}{TeV}$ we find 
\begin{equation}
    W_p \sim 5\times 10^{49}\,(n\,/\,\SI{10}{cm^{-3}})\,(d\,/\,\SI{3.9}{\kpc})^2\,\mathrm{erg}\,.
\end{equation}

As the explanation of the observed \gam-ray emission in a hadronic scenario requires the existence of target material for interactions of the cosmic rays, we infer the presence of hydrogen gas in the vicinity of \wld from radio observations.
We consider both H~I emission (from the SGPS survey \cite{McClureGriffiths2005}) and, as a tracer of dense clouds of molecular hydrogen, CO emission (from \cite{Dame2001}).
We integrate the emission in the velocity interval $(-60,-50)\,\mathrm{km}\,\mathrm{s}^{-1}$, which approximately corresponds to a (near) distance of $\sim\SI{3.9}{\kpc}$, that is, the (supposed) distance of \wld \cite{Kothes2007}.

\begin{wrapfigure}{r}{0.5\textwidth}
    \centering
    \vspace{-0.4cm}
    \includegraphics[width=0.5\textwidth]{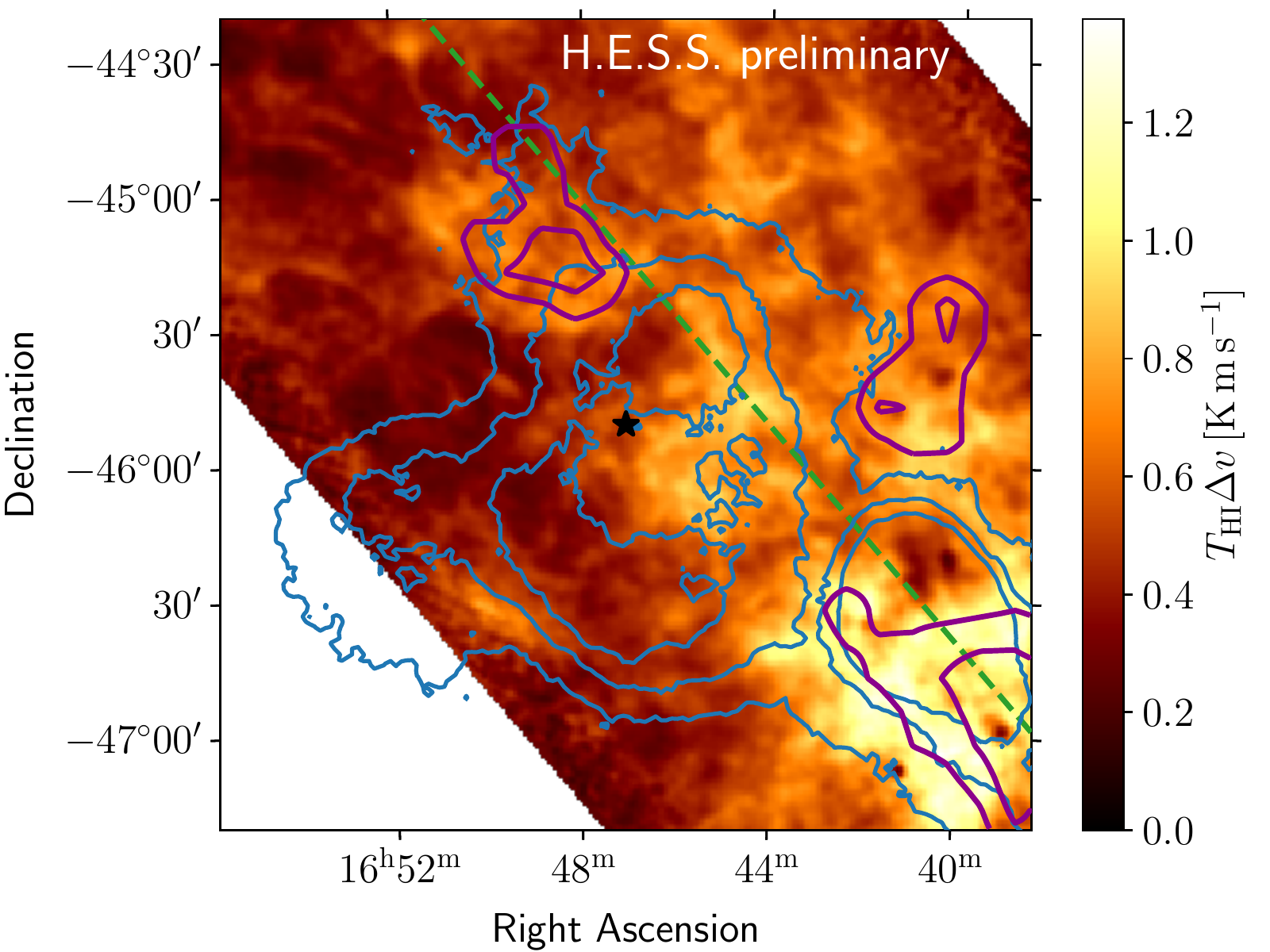}
    \caption{
        Hydrogen gas maps.
        The map in the background shows H~I emission (from \cite{McClureGriffiths2005}) whereas the purple contour lines display CO emission (from \cite{Dame2001}), both integrated in the velocity interval $(-60,-50)\,\mathrm{km}\,\mathrm{s}^{-1}$.
        The blue lines denote significance contours of the \gam-ray emission measured with \hess (same as in Fig.~\ref{fig:sign_map}).
        The black star marks the position of \wld.
    }
    \label{fig:radio}
    \vspace{-1.2cm} 
\end{wrapfigure}

Figure~\ref{fig:radio} shows both the H~I and CO emission, compared with the \gam-ray emission as measured by \hess (this work).
We note that, while the H~I measurements indicate the presence of hydrogen gas in the relevant region, the locations of the dense gas clouds traced by CO do not coincide very well with the regions displaying the highest \gam-ray emission.

\section{Discussion}
With our updated analysis, we confirm many of the findings previously presented in \cite{HESS_Westerlund1_2012}: \hesssrc is a largely extended \gam-ray source exhibiting a very complex morphology.
Its spectrum extends to several tens of TeV and is uniform across the emission region.

We consider first the possibility that (part of) the \gam-ray emission is produced through inverse Compton (IC) scattering of cosmic-ray electrons (``leptonic'' scenario).
Two pulsars with high energy-loss rates ($\dot{E}>\SI{2e35}{\erg\per\s}$), \psrfortyeight and \psrfifty \cite{Manchester2005}, are potential sources of high-energy electrons.
In addition, \wld harbours a magnetar, \cxo, albeit with a lower energy-loss rate ($\dot{E}<\SI{3e33}{\erg\per\s}$) \cite{Muno2006}.
However, taking into account the complex structure of \hesssrc and in particular the lack of energy-dependent morphology (or, equivalently, the uniformity of the energy spectrum across the source region), a scenario in which any combination of these objects is responsible for the entire observed \gam-ray emission is challenging to accomodate.
Notwithstanding this, in particular \psrfortyeight and/or \psrfifty may be contributing to the emission in their immediate surroundings, and we note that we observe a peak in the \gam-ray emission in between the two pulsars.

Turning to hadronic scenarios, in which the \gam-ray emission is produced as a by-product of interactions of cosmic-ray nuclei, we note that no supernova remnant (SNR) -- often favoured as a cosmic-ray source candidate -- has been found in the vicinity of \wld to date.
Nonetheless, the stellar cluster is a plausible source of cosmic rays, which could be accelerated either due to past supernova (SN) activity (up to 150 SNe may have occurred during the cluster life time \cite{Muno2006a}) or in interacting winds of massive stars \cite{Aharonian2019}, or a combination of both.
The observed enhanced \gam-ray emission around \wld may be interpreted as supporting evidence for this, and could indicate that a cavity -- surrounded by a shell of swept-up gas -- has been blown into the interstellar medium by the cluster.
However, the production of \gam rays requires the presence of target material for interactions.
That we do not find an obvious correlation of dense gas clouds as traced by CO emission at $\sim\SI{3.9}{\kpc}$ with the \gam-ray signal may indicate that \wld is located at another distance (see e.g.\ \cite{Aghakhanloo2020}), or that the region harbours gas that is not well traced by CO emission (see e.g.\ \cite{Wolfire2010}).
A more detailed study of possible explanations for the \gam-ray emission observed with \hess will be presented in a forthcoming publication.

\acknowledgments
Full acknowledgements for \hess can be found at \url{https://www.mpi-hd.mpg.de/hfm/HESS/pages/publications/auxiliary/HESS-Acknowledgements-2021.html}.
This research made use of Gammapy\footnote{\url{https://gammapy.org}}, a community-developed core Python package for TeV gamma-ray astronomy \cite{Deil2017}.

\bibliographystyle{JHEP}
\bibliography{bib}

\clearpage
\section*{Full Authors List: \Coll\ Collaboration}

\scriptsize
\noindent
H.~Abdalla$^{1}$, 
F.~Aharonian$^{2,3,4}$, 
F.~Ait~Benkhali$^{3}$, 
E.O.~Ang\"uner$^{5}$, 
C.~Arcaro$^{6}$, 
C.~Armand$^{7}$, 
T.~Armstrong$^{8}$, 
H.~Ashkar$^{9}$, 
M.~Backes$^{1,6}$, 
V.~Baghmanyan$^{10}$
% ...

\medskip

\noindent
$^{1}$University of Namibia, Department of Physics, Private Bag 13301, Windhoek 10005, Namibia\\
$^{2}$Dublin Institute for Advanced Studies, 31 Fitzwilliam Place, Dublin 2, Ireland\\
$^{3}$Max-Planck-Institut f\"ur Kernphysik, P.O. Box 103980, D 69029 Heidelberg, Germany\\
$^{4}$High Energy Astrophysics Laboratory, RAU,  123 Hovsep Emin St  Yerevan 0051, Armenia\\
$^{5}$Aix Marseille Universit\'e, CNRS/IN2P3, CPPM, Marseille, France\\
$^{6}$Centre for Space Research, North-West University, Potchefstroom 2520, South Africa\\
$^{7}$Laboratoire d'Annecy de Physique des Particules, Univ. Grenoble Alpes, Univ. Savoie Mont Blanc, CNRS, LAPP, 74000 Annecy, France\\
$^{8}$University of Oxford, Department of Physics, Denys Wilkinson Building, Keble Road, Oxford OX1 3RH, UK\\
$^{9}$IRFU, CEA, Universit\'e Paris-Saclay, F-91191 Gif-sur-Yvette, France\\
$^{10}$Instytut Fizyki J\c{a}drowej PAN, ul. Radzikowskiego 152, 31-342 Krak{\'o}w, Poland\\
% ...

\end{document}